\documentstyle{kbljrnl}

\begin{document}
\def\G{{\cal G}}
\def\H{{\cal H}}
\def\e{{\rm e}}

%% To be entered at Kluwers: ==>>

%% Uncomment when you want cropmarks to print:
%\printcropmarks

\journame{Journal of Sol--Gel Science and Technology}
\volnumber{??}

%% for VISI, use this form:
%\volnumber{Vol.~no.~2, Issue No.~22}

\issuenumber{??}
\issuemonth{??}
\volyear{1998}

%% For this article:

\received{??}
\revised{??}

%\editor{}
\editors{??}

\authorrunninghead{??}
\titlerunninghead{??}

\setcounter{page}{1} %% This command is optional. 

%% <<== End of commands to be entered at Kluwers 

%%==================================================================
%%  Authors, start here ==>>

%\draft %% Uncomment for double spaces, D R A F T line at bottom of page
          %% which also includes the current date and time of printing.

%% May use \\ to start a new line in title
\title{\begin{center} Phase Behavior of Binary Fluid Mixtures Confined in\\
a Model Aerogel \end{center}}

%% Repeat \author{}\email{}\affil{}
%% for as many authors with separate affiliations as needed.
%% You may also use 
\authors{R. Salazar, R.Toral\thanks{This work is
supported by DGICyT, grants PB94-1167 and PB94-1172 and the 
Agencia Espa\~nola de Cooperaci\'on Internacional in
the Mutis program.}}
\email{rafael@hp1.uib.es,dfsrst0@ps.uib.es}
\affil{Instituto Mediterr\'aneo de Estudios Avanzados (IMEDEA, UIB-CSIC) and\\
Departament de F\'{\i}sica, Universitat de les Illes Balears\\ 
07071 Palma de Mallorca, Spain}

\author{A. Chakrabarti\thanks{This work has been supported by 
the Kansas Center for Advanced Scientific Computing (NSF-EPSCoR) 
and by National Science Foundation grant number DMR-9413513.}}

% Optional: use \email{} for email address, i.e., \email{garcia@cata.spn.edu}
\email{amitc@phys.ksu.edu}

\affil{Department of Physics, Kansas State University, Manhattan, KS
66505, USA}

\begin{abstract}
It is found experimentally that the coexistence region of a
vapor--liquid system or a binary mixture is substantially narrowed when
the fluid is confined in a aerogel with a high degree of porosity
(e.g.  of the order of $95\%$ to $99\%$).  A Hamiltonian model for this
system has recently been introduced \cite{don97}. We have performed
Monte--Carlo simulations for this model to obtain the phase diagram for
the model. We use a periodic fractal structure constructed by
diffusion-limited cluster-cluster aggregation (DLCA) method to simulate
a realistic gel environment. The phase diagram obtained is
qualitatively similar to that observed experimentally.  We also have
observed some metastable branches in the phase diagram which have not
been seen in experiments yet. These branches, however, might be
important in the context of recent theoretical predictions and other
simulations.

\end{abstract}

\keywords{Phase diagram, Aerogels, Monte Carlo simulations, Phase
transitions, surface interaction, confinement effects} % optional

\begin{article}

\section{Introduction}
\label{intro}
When a simple liquid or a binary mixture is confined in a porous
material which has a very low density ($1$--$5\%$) of spatially fixed
impurities, such as in an aerogel, the coexistence region in the phase
diagram is substantially narrowed. This result has been obtained in a
broad class of experimental studies, such as vapor--liquid coexistence
of $^4$He\cite{won90} and Nitrogen\cite{won93}, binary mixtures of
isobutyric acid--water\cite{zhu96} and $^3$He--$^4$He\cite{kim93},
etc.  In all of these studies, the coexistence curve was shown to
change dramatically when the system was confined in a low concentration
silica aerogel.

Recent theoretical efforts have been aimed to understand the above 
mentioned behavior. These include mean--field type studies of
the Random Field Ising model\cite{mar91}, a liquid state approach using
the Ornstein--Zernike equations\cite{pit95} and numerical simulations
of a modified version of the Blume--Emery--Griffiths model\cite{fal95}.
A very successful approach was initiated by Donley and Liu\cite{don97}.
In this reference, the authors introduce a free energy functional that
takes into account the interactions that arise from the contact between
the system molecules and the aerogel.  By minimizing this free energy
they obtain a coexistence curve which is in rough qualitative agreement
with the experimental results. Moreover, the authors go beyond this
mean field type approach by using a parametric form of the equation of
state, combined with linear interpolation techniques. Although this new
approach yields better results than the previous mean field treatment,
it is not conclusive since other parametric models may give different
results. Moreover, as the authors point out correctly, it is very
important to study the role of the fluctuations.

In this paper, we go beyond the mean field approach and numerically
determine the phase diagram of the model introduced in \cite{don97}
by using Monte Carlo methods.  In this model, one considers a scalar
field $m({\bf r})$ and writes down a Hamiltonian which includes bulk
terms plus surface terms coming from the interaction with the aerogel:
\vspace{0.2cm}
\begin{eqnarray} \label{hc}
\H &=& \int_{V} dV \left[ \frac{\theta}{2} m^2({\bf r}) +
\frac{\chi}{4} m^4({\bf r}) \right. \nonumber \\
&& \left. - H m({\bf r}) + \frac 1 2  | \nabla m({\bf r}) |^2 \right]
\nonumber \\
&& + \oint_{S} dS \left[ -H_1 m({\bf r}) +\frac \G 2 m^2({\bf r}) \right]
\end{eqnarray} 
The bulk terms, the first volume integral, is the usual
Ginzburg--Landau model for a scalar concentration field $m({\bf r})$
used in binary phase--transitions. The additional term given by the
surface integral represents the superficial stress\cite{nak83} in the
neighborhood of gel.  Here, the volume $V$ is the available volume for
the fluid and the surface $S$ is the set of fluid points in contact
with the gel.  The parameters for this model are: $\theta$, which is
related to the temperature; $\chi$ which sets the width of the
coexistence curve; the external field (playing the role of the chemical
potential) $H$; the surface field $H_1$; and the surface enhancement
parameter $\G$.

We have performed Monte--Carlo simulations of the lattice version of
the above Hamiltonian in order to find its phase diagram.  We consider
a three-dimensional simple cubic lattice with periodic boundary
conditions. In this lattice, we simulate the presence of the aerogel by
considering that $N_G$ out of the $L^3$ lattice sites belong to a gel
structure generated in a way to be explained in detail later.  We call
these sites ``gel sites".  In the remaining sites (the ``field sites")
we consider the scalar variable $m_i$ ($i=1,\dots,N=L^3-N_G$)
representing the fluid density field.  The gradient term of eq.(1) is
discretized in the usual way:
\begin{equation}
| \nabla m({\bf r}) |^2 \rightarrow \sum_{\mu=1}^3(m_i - m_{i_\mu} )^2
\end{equation}
where $(i_1,i_2,i_3)$ stands for the set of right--nearest--neighbors
sites to site $i$.  However, the presence of the gel has the effect
that in this expression for the gradient: only those neighbor sites
which are actually field sites contribute to the sum.  Accordingly, we
introduce a set of indexes, $O_{i_\mu}$ defined to be equal to 1, when
the site $i_\mu$ is a field site, or 0, when the site $i_\mu$ is a gel
site. The gradient term becomes then:
\begin{equation}
| \nabla m({\bf r}) |^2 \rightarrow \sum_{\mu=1}^3 O_{i_\mu}(m_i - m_{i_\mu} )^2
\end{equation}
For the sake of clarity in notation we have ordered the $N$ field
points such that, from $1$ to $N_B$ we have ``pure bulk" field sites
(i.e. those which are not in contact with the gel) and from $N_B+1$ to
$N$ we have $N_S$ ``surface" field sites, (i.e. those in contact with
the gel, $N_S=N-N_B$). With this convection in mind, the lattice
version of the Hamiltonian (eq.(1)) can be written as:
\vspace{0.2cm}
\begin{eqnarray}
\label{hl}
\H &=& \sum_{i=1}^{N} \left[t m_i^2 + u m_i^4 - h m_i \right. \nonumber \\
&& +\left. \frac 1 2 \sum_{\mu=1}^{3} O_{i_\mu} ( m_i - m_{i_\mu} )^2 
\right] \nonumber \\
&& +\sum_{i=N_B+1}^{N} \left[ -h_1 m_i +g m_i^2 \right]
\end{eqnarray} 

\begin{figure}[t]
\vspace*{6.5cm}
\includegraphics{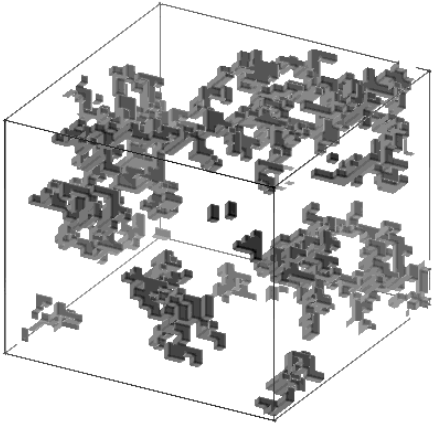}
\caption{\label{fig1}
Three dimensional gel structure with a concentration of $4\%$,
generated by a realization of the DLCA process, in a lattice with $L=32$
and periodic boundary conditions.}
\end{figure}

Where, $t$, $u$, $h$, $h_1$ and $g$ are parameters obtained by suitable
rescaling of the continuum values $\theta$, $\chi$, $H$, $H_1$, $\G$,
respectively.  The gel sites in this lattice form a periodic fractal
structure generated by a diffusion--limited--cluster--aggregation
(DLCA) process\cite{mea83,kol83}, which mimics the aggregation process
that form silica gels.  The algorithm proceeds as follows\cite{has94}:

Let us consider the starting configuration of the gel as a collection of 
aggregates (clusters) containing one particle 
each, the total number of particles 
is $N_G$. At a later time, one obtains a
collection of $N_a$ aggregates, the $i$-th aggregate containing $n_i$ gel
particles, so that
\begin{equation}
\sum_{i=1}^{N_a} n_i = N_G
\end{equation}
The aggregates evolve in the following way:
an aggregate $i$ is chosen at random according to a probability $p_{n_i}$
which depends on the number of particles $n_i$ that it contains, given by
\begin{equation}
p_{n_i}= \frac {n_i^{\alpha}} {\sum_i n_i^{\alpha}}
\end{equation}
with $\alpha=-0.55$.
Then a space direction is chosen at random among the six possible
directions and the cluster is moved by one lattice step in that
direction (we use periodic boundary conditions). If the cluster does
not collide with any other cluster the algorithm continues by choosing
again another cluster at random and moving it. If instead a collision
occurs, the two colliding clusters merge into a new cluster formed by
sticking together the colliding clusters. The process is repeated until
a single cluster remains in the system.  The resulting fractal
dimension of the clusters is $D_F=1.9\pm 0.1$ which is close to the
expected value $D_F=1/\alpha\approx  1.78$.  In Fig.(\ref{fig1}), we
can see a picture of a fractal gel structure obtained using this DLCA
process.

\section{The method}
We use the average value of the field as the order parameter
$\langle M \rangle$:
\begin{equation}
\langle M \rangle =\left\langle \frac{1}{N} \sum_{i=1}^{N} m_i\right\rangle
\end{equation}

\begin{figure*}[t]
\vspace{7.5cm}
\includegraphics{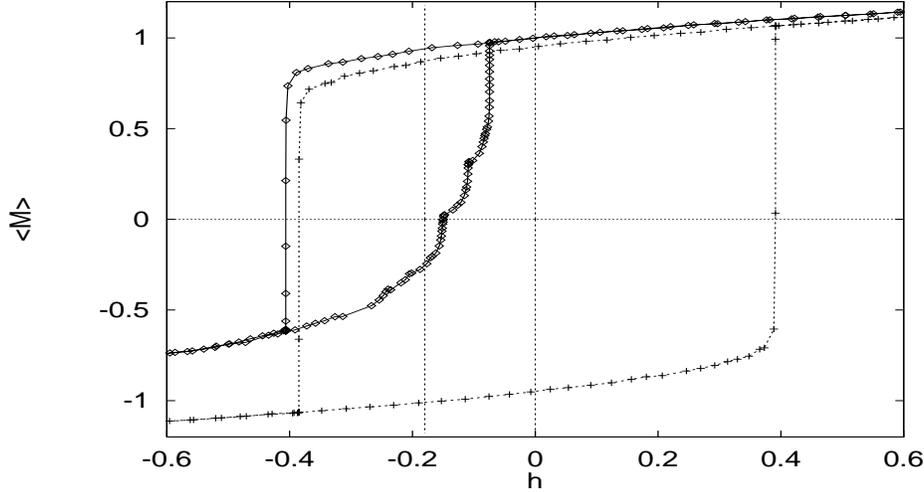}
\caption{\label{fig2}
Hysteresis loops, $\langle M \rangle$ versus $h$, for two cases: The gel
case ($\diamond$) with a concentration of $c=4\%$ and the no gel case ($+$), 
both for a parameter value $t=-1.26$. 
The two vertical lines show the actual value
$h$ where the transition take place, in each case, $h=-0.18$ in
the gel case, and $h=0$ in the no gel case. Note the presence of 
little steps in the lower branch in the gel hysteresis loop.
}
\end{figure*}

where, for fixed gel structure, averages are performed with respect to
the distribution $\e^{-\H}$. For a given gel structure representative
configurations are obtained by the use of the Monte--Carlo method applied
to the lattice Hamiltonian (\ref{hl}). We have used the simple Metropolis
algorithm: a field value $m_i$ is proposed to change to a new
value $m_i'$ chosen randomly from a uniform distribution in $(m_i-\delta,
m_i+\delta)$ for given $\delta$. The new value $m_i'$ is accepted with a 
probability given by $\min[1,\e^{-\Delta\H}]$, with $\Delta\H= \H'-\H$
is the change in the Hamiltonian implied by the proposed change.
The order parameter $\langle M \rangle$ is computed as an average over
different field configurations. An additional average has been performed with
respect to 10 different gel structures. 

To find the phase diagram, i.e. 
the dependence on the ``temperature" $t$ of 
the order parameter $\langle M \rangle$, we take 
fixed values for the system parameters $u$, $g$ and $h_1$, 
and vary the ``temperature" $t$. For each value of the 
temperature $t$ we compute the hysteresis loop by using the Monte--Carlo
method varying 
the external field $h$ from $+h^0$ to $-h^0$ and vice-versa. 
We first start at a sufficiently high value for $h^0$ (see later)
and compute $\langle M \rangle^0$. Next, by keeping the
same final configuration for the density field, 
the external field is lowered
by an amount $\Delta^0$ to $h^1=h^0-\Delta^0$ and compute the
corresponding value for the order parameter $\langle M \rangle^1$. 
Then the field is changed to $h^2=h^1-\Delta^1$ and 
so on until we arrive at $-h^0$. The process is reversed by
increasing in a similar way the external field
to reach again $+h^0$. 

In order to determine accurately the hysteresis loop, 
we do not take a constant value for $\Delta^i$ but we take:
\begin{equation}
\Delta^i=\Delta^0 \frac {\Delta^{i-1}} 
{\sqrt{(h^i-h^{i-1})^2+(\langle M \rangle^i-\langle M \rangle^{i-1})^2
\alpha^2}}
\end{equation}
where $\alpha$ is an additional scale control parameter.
This means that we control the length along the 
hysteresis curve allowing us to have smooth hysteresis curves.
Two typical results for the hysteresis loops are shown in Fig.(\ref{fig2})
for the cases of no gel and a gel filling 4\% of the lattice points.
In the no--gel case, the hysteresis loop is symmetrical around $h=0$ and
one can read directly the equilibrium values for $\pm \langle M \rangle$ by
taking the values at $h=0$. When the gel is present, we determine the
equilibrium values for $\langle M \rangle$ by demanding that the
Gibbs free energy in the two phases is equal. 
The Gibbs free energy can be obtained by integration of the general 
relation\cite{hua87}:
\begin{equation}
\langle M \rangle=-\frac {\partial G}{\partial h} 
\end{equation}
By integrating along the upper curve of the hysteresis loop we obtain:
\begin{equation}\label{gfe}
G^{(1)}(h)=G(h^0)-\int_{h^0}^h \langle M \rangle dh
\end{equation}
whereas from the lower part of the hysteresis loop:
\begin{equation}\label{gfei}
G^{(2)}(h)=G(-h^0)-\int_{-h^0}^h \langle M \rangle dh
\end{equation}

The equilibrium values for $\langle M \rangle$ are read from the
hysteresis loop at the value of the external field $h$ such that
$G^{(1)}(h)=G^{(2)}(h)$. In order to compute those values for the free
energy, according to (\ref{gfe}) and (\ref{gfei}) we need to know
the values of $G(h^0)$ and $G(-h^0)$. For this, we use a sufficiently
large value for $h^0$.
For such a large external field, the mean 
field is a good approximation, in such a way that the Gibbs free--energy
can be replaced by just the internal energy $\H$. So  
we take $G(\pm h^0) \approx \H(\pm h^0)$. We have taken $h^0=10$.
In order to check the validity of mean field for this value of $h^0$ we
have compared the resulting average $\langle M \rangle^0$ obtained in
the simulation with the mean field value obtained from minimizing the
Hamiltonian $\H$ for the same value for $h^0$. Both results agree
within 1\%.

\begin{figure*}[t]
\vspace{7.5cm}
%\vskip1in
\includegraphics{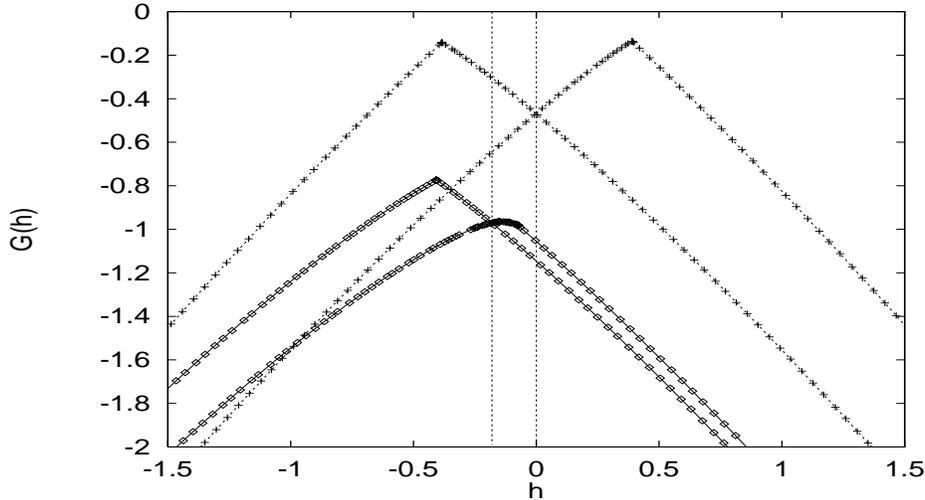}
\caption{\label{fig3}
The Gibbs free--energy $G(h)$ versus $h$ obtained by integration, Eqs.
(\ref{gfe},\ref{gfei}), from the hysteresis curves in Fig.(\ref{fig2}), in the
gel case ($\diamond$) and the no gel case ($+$). From these curves we deduce the
necessary value for the external field $h$, where  the first-order
transition takes place.
}
\end{figure*}

In Fig.(\ref{fig3}) we plot the results of the numerical integrations
(\ref{gfe}) and (\ref{gfei}) both for the no-gel case and for one gel
configuration of porosity 96\% for a value of the parameter 
$t=-1.26$. As expected, in the no gel case, $G^{(1)}(h)$ and $G^{(2)}(h)$
coincide for $h=0$. In the gel case, we read from this curve the 
corresponding value for $h\approx -0.18$. Using this value, we 
obtain from the upper and lower curves of the hysteresis loops,
see Fig.(\ref{fig2}), the corresponding values for $\langle M \rangle$.

We have found that this method can be used efficiently 
far enough from the critical point. Near the critical point, the numerical
errors produce a large uncertainty in the numerical integrations and it is
difficult to accurately determine the required value of the external field.
In those cases, we have taken simply an average of the lower and upper branches
of the hysteresis loop as the values for $\langle M \rangle$. 
For temperatures above the critical one, there is no hysteresis loop.

\section{Phase Diagram}
We present in this section the phase diagram as a function of the parameter
$t$ for three different cases: (i) the no--gel situation,
(ii) a gel case with a porosity of 88\% and (iii) a gel case with a porosity
of 96\%.
\begin{figure*}[t]
\vspace{7.5cm}
\includegraphics{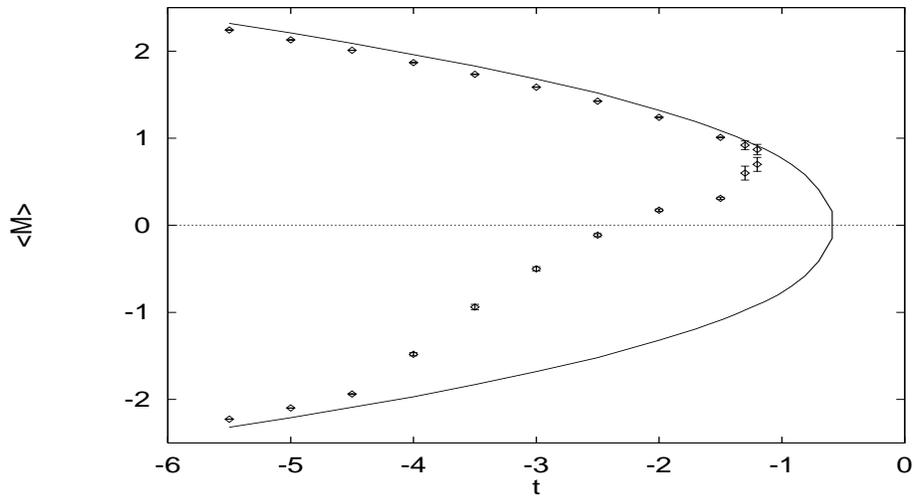}
\caption{\label{fig4}
Phase diagrams, in the gel ($\diamond$), $c=12\%$, and no gel 
(continuous line) cases. 
}
\end{figure*}
We use in all the cases a lattice with $L^3=32^3$ sites and   
the common Hamiltonian parameter values
$u=0.5$, $h_1=4$ and $g=1$. For the factors controlling the step size
for the variation of the external field we take 
$\alpha=3$, $\Delta^0=0.05$. 
The initial values for  the hysteresis loop is $h^0=10$.
In the gel cases, we have taken averages with respect to 10 gel structures.
By following the method described in the previous
section, we obtain for each temperature two values for the 
order parameter $\langle M \rangle$. These are plotted in Fig.(\ref{fig4})
and Fig.(\ref{fig5}) for a porosity of 88\% and 96\%, respectively.
In these figures, we can see clearly 
the narrowing of the coexistence region when 
the gel is present. For smaller porosity (larger fraction of the gel)
the narrowing is more pronounced as observed in the experiments and
in accordance with the results of mean field theory. 
Although, as mentioned at the end of the previous section, numerical
errors become large near the critical point, we conclude from the
figures that 
the critical temperature is lowered when the gel is present. Again, the
reduction in the critical temperature is larger for smaller porosity.
\begin{figure*}[b]
\vspace{7.5cm}
\includegraphics{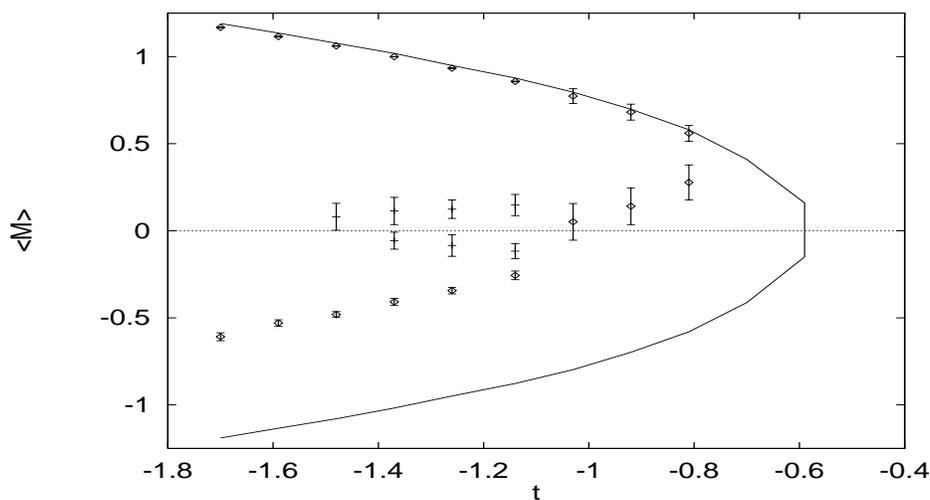}
\caption{\label{fig5}
Phase diagrams, in the gel ($\diamond$), $c=4\%$, and no gel (continuous line) cases. 
Note in the gel case that we have included some points ($+$), corresponding to 
the steps found in the hysteresis loops. 
}
\end{figure*}
However, we note that in the simulations where the gel is present, we
have found some steps in the lower curve of the hysteresis loop, as we
can see for instance in Fig.(\ref{fig2}), which could be interpreted as
signaling a second transition.  We have plotted in the phase diagram
additional points corresponding to the steps in the hysteresis loops.
The location of those points depends strongly on the particular
realization of the DLCA process to generate the fractal gel structure.
This shows up in the large error bars for these points in the
Fig.(\ref{fig5}).

The phase diagram obtained in this paper is qualitatively similar to
that observed experimentally\cite{won90}:  the coexistence region in
presence of gel is narrowed and shifted  with respect to the non--gel
situation.  There are marks of a second transition which also show up
in the mean field studies of reference (\cite{don97}) and also in other
simulations of the Lennard--Jones fluid\cite{pag96}, although it has
not been reported in experimental studies.

\acknowledgements
We wish to thank T. Sintes for several discussions about 
DLCA methods. A.C. thanks Andrea Liu for many useful discussions and for
sharing unpublished results with us.
%\appendix{}

%\notes % You need \notes only if you don't use references.

%% \bio{Author Name} Text of biography...
%% Repeat as many times as necessary.
%\bio{1in}

\end{article}
\end{document}